\documentclass{iaus}
\usepackage{graphicx}
\title[Magnetic structure of our Galaxy] 
{Magnetic structure of our Galaxy:\\ A review of observations}

\author[J. L. Han]   
{JinLin Han}
\affiliation{National Astronomical Observatories, Chinese Academy of
  Sciences \break Jia-20 DaTun Road, Chaoyang District, Beijing 100012,
China \break email: hjl@bao.ac.cn}

\pubyear{2009}
\volume{259}  
\pagerange{455--466}
\date{2008 Dec. 10th and in revised form 2008 Dec. 25th}
 \jname{Cosmic Magnetic Fields: From Planets,
to Stars and Galaxies} \editors{K.G. Strassmeier, A.G. Kosovichev \&
J.E. Beckman, eds.}

\begin{document}

\maketitle

\begin{abstract}
The magnetic structure in the Galactic disk, the Galactic center and the
Galactic halo can be delineated more clearly than ever before.
In the Galactic disk, the magnetic structure has been revealed by starlight
polarization within 2 or 3 kpc of the Solar vicinity, by the distribution of
the Zeeman splitting of OH masers in two or three nearby spiral arms, and by
pulsar dispersion measures and rotation measures in nearly half of the
disk. The polarized thermal dust emission of clouds at infrared, mm and
submm wavelengths and the diffuse synchrotron emission are also related to
the large-scale magnetic field in the disk. The rotation measures of
extragalactic radio sources at low Galactic latitudes can be modeled by
electron distributions and large-scale magnetic fields. The statistical
properties of the magnetized interstellar medium at various scales have been
studied using rotation measure data and polarization data.
In the Galactic center, the non-thermal filaments indicate poloidal
fields. There is no consensus on the field strength, maybe mG, maybe tens of
$\mu$G. The polarized dust emission and much enhanced rotation measures of
background radio sources are probably related to toroidal fields.
In the Galactic halo, the antisymmetric RM sky reveals large-scale toroidal
fields with reversed directions above and below the Galactic plane.
Magnetic fields from all parts of our Galaxy are connected to form a global
field structure. More observations are needed to explore the untouched
regions and delineate how fields in different parts are connected.
\end{abstract}

\firstsection 
\section{Introduction}

Radio observations measure the total intensity, polarized intensity and
rotation measure (RM). Sky maps of these quantities are all related to the
magnetic fields in our Galaxy. Because the Milky Way is the largest edge-on
galaxy in the sky, the synchrotron emission from relativistic electrons
gyrating in the Galactic magnetic field is dominant in the radio sky. The
stronger radio emission is observed nearer the Galactic plane, and strongest
towards the Galactic central regions. The Galactic radio emission is
polarized (\cite{rei07}), if it is not depolarized, best shown at tens of
GHz (\cite{phk+07}). The polarized emission undergoes Faraday rotation due
to the Galactic magnetic fields and ionized electrons. The Faraday sky,
i.e. the RM sky, is strikingly antisymmetric to the Galactic coordinate
(\cite{hmbb97}, see Fig.~\ref{rmsky}).

The magnetic field of our Galaxy is more important in astrophysics and
astroparticle physics than the fields in other galaxies.
Magnetic fields are certainly one of key ingredients of the interstellar
medium.
Large-scale magnetic fields contribute to the hydrostatic balance and
stability of the interstellar medium (\cite{bc90}), and even disk dynamics
(\cite{bf07}).
Magnetic fields in molecular clouds, which are closely related to the
Galactic fields (\cite{hz07}), play an important role in the star
formation process (see a review by \cite{hc05}).
More important is that the magnetic fields of our Galaxy are the main 
agent for transport of charged cosmic-rays (e.g. \cite{tt02,ps03}).  It is
impossible to understand the origin and propagation of cosmic rays without
adequate knowledge of Galactic magnetic fields.
The Galactic radio emission and its polarization, which result from the
Galactic magnetic fields, is found to heavily (up to 95\% in polarization)
``pollute'' the measurements of the cosmological microwave background (CMB,
e.g. \cite{phk+07}). Galactic magnetic fields have suddenly become very
important in the CMB studies of cosmology!

To understand magnetic fields, we have first to measure them and learn their
properties. The galactic scale is in the middle between the stellar scale
and the cosmological scale. It is on this scale that the magnetic fields can
be well {\it measured} at present, at least from our Galaxy.  Observations
of galactic-scale magnetic fields provide the most important hints for and
constraints on the origin and maintenance of the magnetic fields of Galactic
objects and the fields in the universe. In the following, I will review the
observed magnetic structures in the Galactic disk, the Galactic center and
the Galactic halo.

\section{Magnetic fields in the Galactic disk}

Magnetic fields pervade the diffuse interstellar medium, molecular clouds,
and very dense cloud cores or HII regions. When interstellar gas contracts
to form a cloud or a cloud core the field is enhanced. Therefore, the
observed field strength in clouds increases with gas density
(\cite{cru99}). Here, I review the observational results of large-scale
magnetic fields in the Galactic disk which are related to the diffuse medium
and spiral structure and have a scale greater than 1~kpc. Magnetic fields on
smaller scales will be mentioned only if they are related to the large-scale
fields.

There are five observational tracers of the Galactic magnetic fields:
polarization of starlight, polarized thermal dust emission from clouds,
Zeeman splitting of lines, diffuse synchrotron radio emission, and Faraday
rotation of polarized sources.

\noindent{\bf Polarization of starlight}

Starlight becomes polarized when it passes through the interstellar
medium and is absorbed or scattered by interstellar dust grains
preferentially aligned by magnetic fields. The observed polarization
is the integrated effect of scattering between the star and the Sun,
and the ``polarization vectors'' show the averaged field orientation
(weighted by the unknown local dust content). Starlight polarization
data have been obtained for about 10,000 stars, mostly within 2 or
3~kpc of the Sun (see \cite{mf70,hei00}). Analysis of these data show
that the local field is parallel to the Galactic plane and follows the
local spiral arms (\cite{hei96}). Starlight polarization data are
difficult to use for detection of Galactic magnetic fields in a much
larger region than 2 or 3 kpc. However, recent developments in
instruments help to get a lot of new starlight polarization data
for revealing magnetic fields in given objects (e.g. \cite{fvmo08}).

\noindent{\bf Polarized thermal dust emission from clouds}

In recent years, with development of instruments and backend
technology (e.g. \cite{hdd+00}), observations of polarized thermal
dust emission at mm, sub-mm and infrared bands have been used to detect the
transverse orientation of magnetic fields in molecular clouds (see
review by \cite{hc05}) on scales from 1 pc to tens of pc. The
observed magnetic fields always have an hourglass shape, which
indicates that the fields were enhanced when the clouds were formed by
compressing the diffuse interstellar medium.  Recently, Li et
al. (2006)\nocite{lgk+06} found that magnetic fields in molecular
clouds seem to be preferentially parallel to the Galactic plane,
indicating that the magnetic fields in the clouds preserve the fields
frozen into the diffuse medium.

\noindent{\bf Zeeman splitting of lines}

Zeeman splitting of spectral lines can measure {\it in situ} field
strength of the line-of-sight component in molecular clouds or maser spots
with a scale size $<1$ AU.
To date, Zeeman splitting of emission or absorption lines has been
detected from about 20 clouds (see a list in \cite{cru99,hz07}) and in
OH masers associated with 140 HII or star-formation regions using
single dishes (e.g. \cite{c03}) or interferometers
(e.g. \cite{fram03}).
From collected Zeeman splitting data of OH masers of HII regions and
OH or HI absorption or emission lines of molecular clouds themselves,
Han \& Zhang (2007)\nocite{hz07} found large-scale reversals in the
sign of the line-of-sight component of the median field, indicating
field reversals in a pattern similar to reversals revealed from pulsar
RM data (see below). Evidently, magnetic fields on such a small scale
are related to the large-scale Galactic magnetic fields to a very
surprising extent (\cite{rs90,fram03}). Interstellar magnetic fields
are apparently preserved as fossil fields through the cloud formation
and star formation process, and even in stars (see G. Wade's talk in
this volume). How can such coherent magnetic field directions be
preserved from the low density medium ($\sim1 {\rm cm}^{-3}$) to
higher density clouds ($\sim10^{3}{\rm cm}^{-3}$), even to the highest
density in maser regions ($\sim10^{7}{\rm cm}^{-3}$), with compression
of 3 or even 10 orders of magnitude? It is a puzzle. The turbulent and
violent processes in molecular clouds apparently cannot significantly
alter the mean magnetic field.

\noindent{\bf Diffuse synchrotron radio emission}

The total and polarized intensities of synchrotron emission are
usually used to estimate the total and {\it ordered} field
strength. For nearby galaxies, the strength of the {\it regular}
large-scale field is probably much overestimated from the polarization
percentage, because the so-called {\it ordered} fields consist of the
regular large-scale fields and anisotropic random fields
(\cite{hbe+99}) which both produce polarized emission.

We cannot have a face-on view of the global magnetic field structure
in our Galaxy through polarized synchrotron emission, as is possible
for nearby spiral galaxies (see R. Beck's talk in this volume).
Polarization surveys of the Galactic plane have been comprehensively
reviewed by Reich (2007)\nocite{rei07}. The observed polarized
emission from the Galactic plane is the sum of all contributions with
different polarization properties (i.e. polarization angles and
polarization percentages) coming from various regions at different
distances along a line of sight. Emission from more distant regions
suffers from more Faraday effect produced by the foreground
interstellar medium. Polarized emissions from different
regions should ``depolarize'' each other when they are summed, which
is more obvious at lower frequencies. Observations at higher
frequencies (e.g.  at 6~cm by \cite{shr+07}) show polarized structures
at larger distances because depolarization is less severe. The
polarized structures are closely related to the magnetic field
structure where the emission is generated. Some large-angular-scale
polarized features are seen emerging from the Galactic disk, for
example, the North Polar Spur (e.g. \cite{jfr87}).

\noindent{\bf Faraday rotation of pulsars and radio sources}

Faraday rotation of linearly polarized radiation from pulsars and
extragalactic radio sources (EGRs) is the most powerful probe of the diffuse
magnetic field in the Galaxy (e.g. \cite{hml+06,bhg+07}). Magnetic fields in
a large part of the Galactic disk have been revealed by RM data of pulsars,
which gives a measure of the line-of-sight component of the field. EGRs have
the advantage of large numbers but pulsars have the advantage of being
spread through the Galaxy at approximately known distances, allowing direct
three-dimensional mapping of the magnetic fields. For a pulsar at distance
$D$ (in pc), the RM (in radians~m$^{-2}$) is given by
$ 
{\rm RM} = 0.810 \int_{0}^{D} n_e {\bf B} \cdot d{\bf l},
$ 
where $n_e$ is the electron density in cm$^{-3}$, ${\bf B}$ is the vector
magnetic field in $\mu$G and $d {\bf l}$ is an elemental vector along the
line of sight toward us (positive RMs correspond to fields directed toward
us) in pc. With the pulsar dispersion measure,
$ 
{\rm DM}=\int_{0}^{D} n_e d l,
$ 
we obtain a direct estimate of the field strength weighted by the local free
electron density
\begin{equation}\label{eq_B}
\langle B_{||} \rangle  = \frac{\int_{0}^{D} n_e {\bf B} \cdot d{\bf
l} }{\int_{0}^{D} n_e d l } = 1.232 \;  \frac{\rm RM}{\rm DM}.
\label{eq-B}
\end{equation}
where RM and DM are in their usual units of rad m$^{-2}$ and cm$^{-3}$ pc
and $B_{||}$ is in $\mu$G.

\begin{figure*}[bt]
\begin{center}
\includegraphics[angle=270,width=0.96\textwidth]{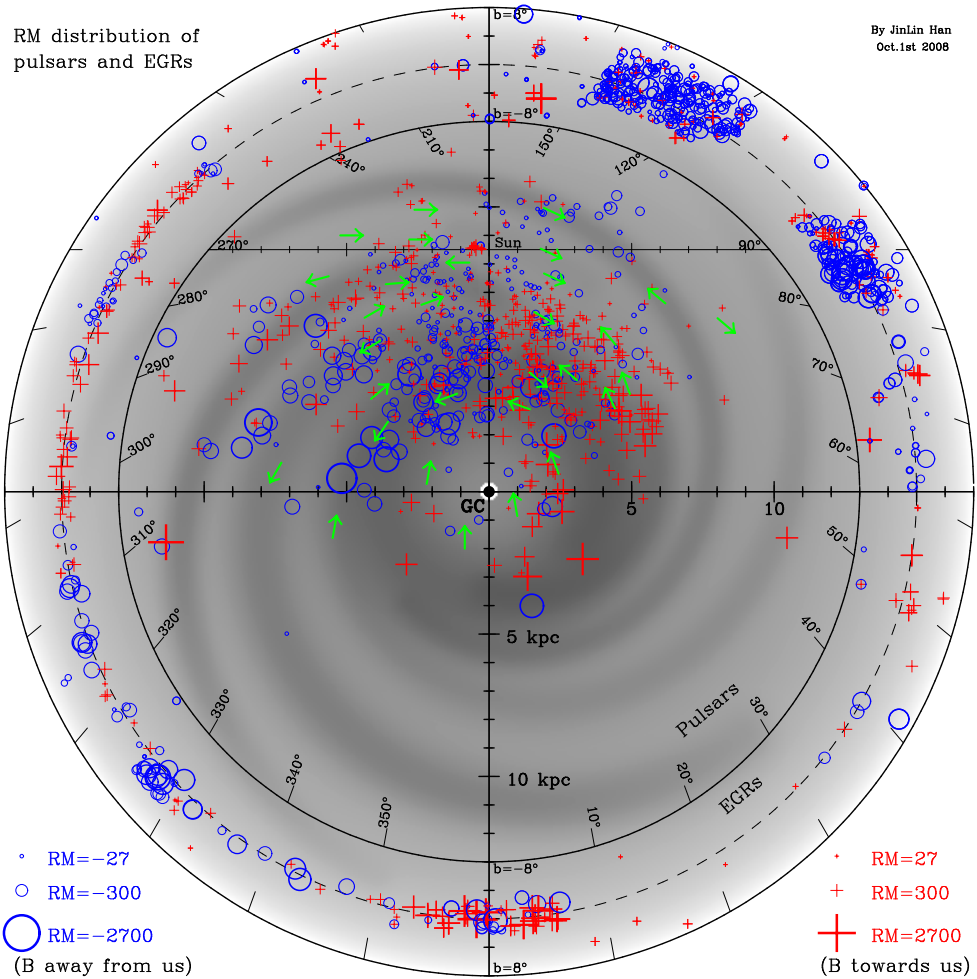}
\caption{The RM distribution of 736 pulsars with $|b|<8^{\circ}$ projected
onto the Galactic plane, including new data of Han et al (2009, in
preparation). The linear sizes of the symbols are proportional to the square
root of the RM values with limits of $\pm$27 and $\pm$2700
rad~m$^{-2}$. Positive RMs are shown by plus signs and negative RMs by open
circles. The background shows the approximate locations of spiral arms used
in the NE2001 electron density model (\cite{cl02}). RMs of 1285 EGRs of
$|b|<8^{\circ}$ (data mainly from \cite{ccsk92,gdm+01,btj03,rrs05,bhg+07}
and other RM catalogs) are displayed in the outskirt ring according to their
$l$ and $b$, with the same convention of RM symbols and limits. The
large-scale structure of magnetic fields indicated by arrows was derived
from pulsar RMs and comparison of them with RMs of background EGRs (details
in Han et al. 2009). The averaged RM fluctuations with Galactic longitudes
of EGRs are consistent with magnetic field directions derived from pulsar
data, for example, in the 4th Galactic quadrant. \label{rm_psr_egr}}
\end{center}
\vspace{-3mm}
\end{figure*}

Previous analyses of pulsar RM data have often used the model-fitting method
(\cite{man74,tn80,rk89,hq94,id99,njkk08}), i.e., to model magnetic field structures in
all of the paths from the pulsars to us (observer) and fit them, together
with the electron density model, to the observed RM data. {\it Significant
improvement} can be obtained when both RM and DM data are available for many
pulsars in a given region with similar lines of sight. Measuring the
gradient of RM with distance or DM is the most powerful method of
determining both the direction and magnitude of the large-scale field local
to that particular region of the Galaxy
(\cite{ls89,hmq99,hmlq02,wck+04,hml+06}). Field strengths in the region can
be {\it directly derived} from the slope of trends in plots of RM versus
DM. Based on Eq.~(\ref{eq-B}), we get

\begin{equation}
\langle B_{||}\rangle_{d1-d0} = 1.232 \frac{\Delta{\rm RM}}{\Delta{\rm DM}}
\label{delta_rm_dm}
\end{equation}
where $\langle B_{||}\rangle_{d1-d0}$ is the mean line-of-sight field
component in $\mu$G for the region between distances $d0$ and $d1$,
$\Delta{\rm RM} = {\rm
RM}_{d1} - {\rm RM}_{d0}$ and $\Delta{\rm DM} ={\rm DM}_{d1} - {\rm
DM}_{d0}$.

\vspace{1mm}\noindent{\bf 1). Field structure:}

So far, RMs of 1021 pulsars have been observed
(\cite{hl87,rl94,qmlg95,hmq99,wck+04,hml+06,njkk08}), if new RMs of 477 pulsars
by Han, van Straten, Manchester \& Demorest (2009, in preparation) are
included (see Fig.~\ref{rm_psr_egr}). This enables us to investigate the
structure of the Galactic magnetic field over a larger region than that
previously possible. We have detected counterclockwise magnetic fields in
the innermost arm, the Norma arm (\cite{hmlq02}). A more complete analysis
for the fields of our Galaxy (\cite{hml+06} and Han et al. 2009, in
preparation) from both RMs of pulsars and EGRs gives a picture for the
coherent large-scale fields aligned with the spiral-arm structure, as shown
in Fig.1: magnetic fields in all inner spiral arms are counterclockwise when
viewed from the North Galactic pole. On the other hand, at least in the
local region and in the inner Galaxy in the fourth quadrant, there is good
evidence that the fields in interarm regions are similarly coherent, but
reversed to be clockwise. There are at least four or five reversals in the
fourth quadrant, probably occurring near the boundary of the spiral arms. In
the Galactic central region interior to the Norma arm, new RM data of
pulsars indicate that the fields are clockwise, reversed again from the
counterclockwise field in the Norma arm. In the first Galactic quadrant,
because the separations between spiral arms are so small, the RM data are
dominated by counterclockwise fields in the arm regions though some (not
many) negative pulsar RMs indicate clockwise fields in the interarm
regions. The magnetic field in the Perseus arm cannot be determined well.

The averaged RM of EGRs in a given sky region reflect the common foreground
Galactic RM contribution, which is the integration of $n_eB$ from the Sun to
the outskirts of the Galactic disk. Comparison of the mean of RMs of
background EGRs with RMs of foreground pulsars can reveal the magnetic
fields behind the pulsars (see Fig.~\ref{rm_psr_egr}). However, the dominant
contribution of RMs of EGRs behind the Galactic disk comes from the
interstellar medium mostly in tangential regions. The fluctuations in the RM
distribution of extragalactic radio sources
(\cite{ccsk92,gdm+01,btj03,rrs05,bhg+07}) with Galactic longitude,
especially these of the fourth Galactic quadrant, are consistent with
magnetic field directions derived from pulsar data (see Fig.~1). The
negative RMs of EGRs in the 2nd quadrant suggest that the interarm fields
both between the Sagittarius and Perseus arms and beyond the Perseus arm are
predominantly clockwise.

\begin{figure}[b]
  \begin{minipage}[t]{0.47\linewidth}
    \centering \includegraphics[angle=270,width=62mm]{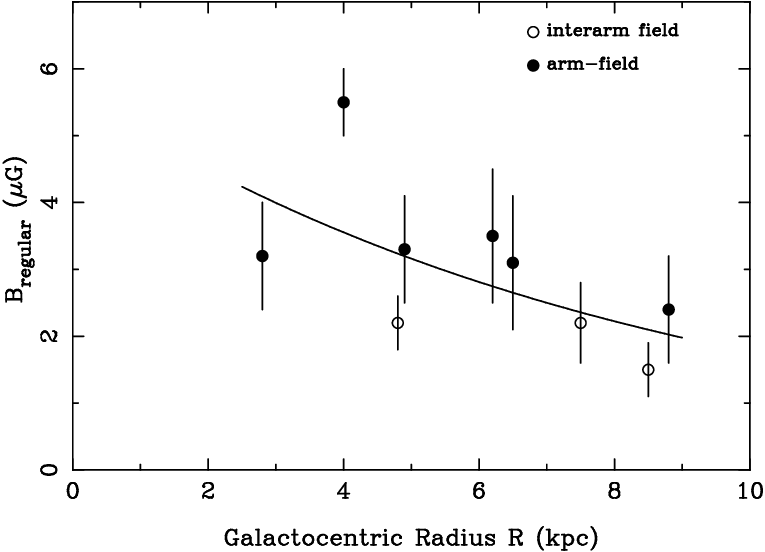}
    \caption{\small Variation of the large-scale regular field strength with
      Galactocentric radius derived from pulsar RM and DM data near the
      tangential regions (Han et al. 2006). Note that the ``error-bars'' are
      not caused by the uncertainty of the pulsar RM or DM data, but reflect
      the random magnetic fields in the regions.\label{br}}
  \end{minipage}%
  \hspace{0.02\textwidth}
  \begin{minipage}[t]{0.51\textwidth}
    \centering
    \includegraphics[width=48mm,height=68mm,angle=-90]{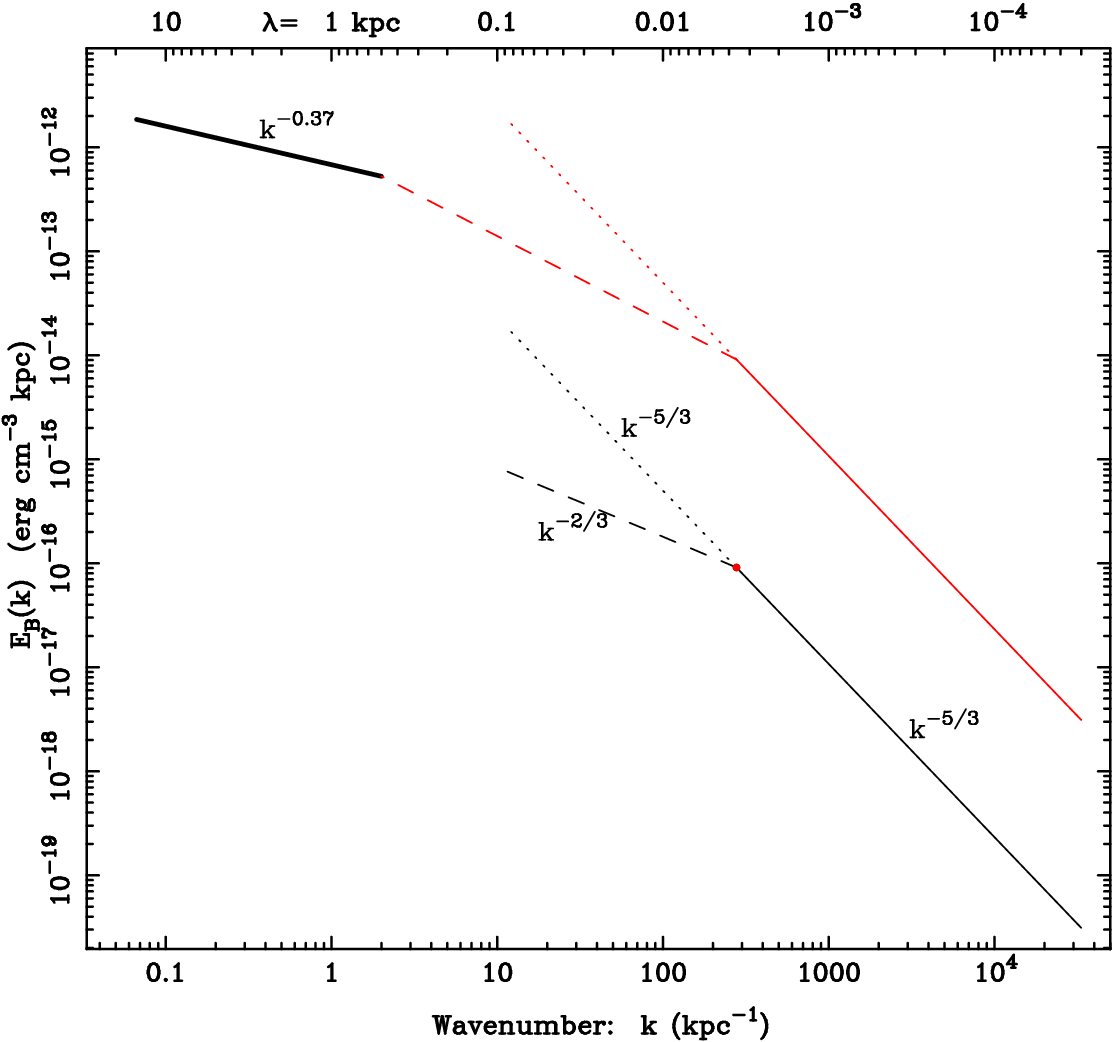}
    \caption{\small Composite spatial magnetic-energy spectrum in our
      Galaxy.  See Han et al. (2004) for details.  The thin solid and
      dashed/dotted lines at smaller scales are the Kolmogorov and
      2D-turbulence spectra derived from Minter \& Spangler (1996), and the
      upper ones from new measurements of Minter (2004, private
      communication).\label{eb}}
  \end{minipage}%
  %
\end{figure}

Fitting various models to the new RM dataset shows different results
(e.g. \cite{njkk08}). The most important is to have a model consistent with
the detected field reversals. Looking at Table 2 of Men et
al. (2008)\nocite{mfh08}, one can see that any single-mode model is not
enough to fit RM data though the bisymmetrical spiral model with field
reversals is the best and gives the smallest $\chi^2$.

\vspace{1mm}\noindent{\bf 2). Field strength:}

With much more pulsar RM data, for the first time, Han et al. (2006)
\nocite{hml+06} were able to {\it measure} the regular azimuthal field
strength of near the tangential regions in the 1st and 4th Galactic
quadrants, and then plot the dependence of regular field strength as a
function of galactocentric radius (Fig.~\ref{br}). Although the
``uncertainties'', which in fact reflect the random fields, are large, the
tendency is clear that fields get stronger at smaller Galactocentric radius
and weaker in interarm regions. To parameterize the radial variation, an
exponential function was chosen to give the smallest $\chi^2$
value and to avoid a singularity at $R=0$ (for $1/R$) and unphysical values
at large R (for the linear gradient). That is, $ B_{\rm reg}(R) =
B_0 \; \exp \left[ \frac{-(R-R_{\odot})} {R_{\rm B}} \right] , $ with the
strength of the large-scale field at the Sun, $B_0=2.1\pm0.3$ $\mu$G, and
the scale radius $R_{\rm B}=8.5\pm4.7$ kpc.

In addition, the vertical field in the solar vicinity $B_z$ is estimated to
be $\sim 0.2 \mu$G (\cite{hq94,hmq99}). Using RMs of EGRs, Simard-Normandin
\& Kronberg (1980) and Han \& Qiao (1994) \nocite{sk80,hq94} estimated the
scale-height of Galactic magnetoionic disk about 1.4~kpc.

\vspace{2mm}\noindent{\bf 3). Field statistics on small scales and spatial
  B-energy spectrum:}

Magnetic fields in our Galaxy exist on all scales.  For the large-scale
field, we can determine the field structure and field strength. To study
small-scale magnetic fields, the only approach is to make statistics and
have a description of their statistical properties.

Pulsar RMs have been used to study the small-scale random magnetic fields in
the Galaxy. Some pairs of pulsars close in sky position have similar DMs but
very different RMs, indicating an irregular field structure on scales of about
100~pc (\cite{ls89}). Some of these irregularities may result from HII
regions in the line of sight to a pulsar (\cite{mwkj03}).  It has been found
from pulsar RMs that the random field has a strength of $B_r\sim 4-6\mu$G
independent of cell-size in the scale range of 10 -- 100~pc (\cite{rk89},
\cite{os93}). From pulsar RMs in a very large region of the Galactic disk,
Han et al. (2004)\nocite{hfm04} obtained a power law distribution for
magnetic field fluctuations of $E_B(k)= C \ (k / {\rm
kpc^{-1}})^{-0.37\pm0.10}$ at scales from $1/k=$ 0.5~kpc to 15~kpc, with $C=
(6.8\pm0.8)\ 10^{-13} {\rm erg \ cm^{-3} \ kpc}$, corresponding to an rms
field of $\sim6\mu$G in the scale range.

Some results of magnetic field statistics were also obtained from RMs of EGRs.
Armstrong et al. (1995)\nocite{ars95} showed that the spatial power spectrum
of electron density fluctuations from small scales up to a few pc could be
approximated by a single power law with a 3D spectral index $-$3.7, very
close to the Kolmogorov spectrum. Magnetic fields on such small scales
should follow the same power-law, since magnetic fields are frozen into
the ionized interstellar medium.
Minter \& Spangler (1996)\nocite{ms96} found that structure functions of
rotation measure and emission measure were consistent with a 3D-turbulence
Kolmogorov spectrum of magnetic fields up to 4~pc, but with a 2D turbulence
between 4~pc and 80 pc.
Haverkorn et al. (2006)\nocite{hgb+06} found that the structure
functions of RMs in the directions tangential to the arms have much
larger slopes than those in the interarm directions, indicating that
the arm regions are more turbulent. 
Sun \& Han (2004)\nocite{sh04} found that the structure function for RMs at
the two Galactic poles is flat, but at lower latitudes it becomes inclined
with different slopes at different Galactic longitudes.

Combining the above information, one can get the spatial energy
spectrum of Galactic magnetic fields (see \cite{hfm04}) which should
constrain the theoretical simulations (e.g. \cite{bk05}) of the
generation and maintenance of Galactic magnetic fields. Evidently, on
small scales, the distribution follows the Kolmogorov power-law
spectrum, but at larger scales, it becomes flat and probably has
two breaks at a scale of few pc and several tens of pc
(Fig.~\ref{eb}). The interstellar magnetic fields probably become
strongest at the scales of energy injection by supernova explosions
and stellar winds (e.g. 1 to 10~pc).

\begin{figure}[b]
\includegraphics[width=135mm]{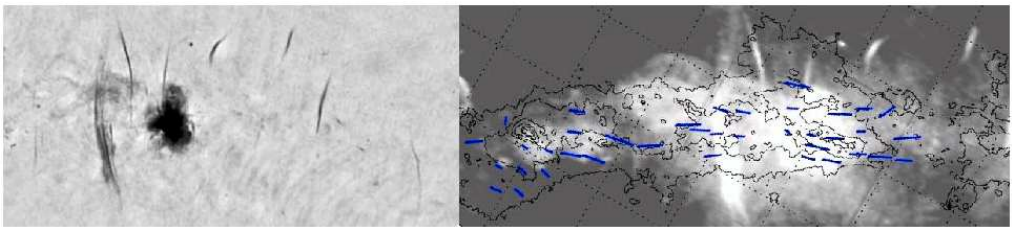}
\caption{Non-thermal radio filaments discovered in the Galactic center
($\lambda$90cm data from \cite{lnlk04}) and the polarized thermal dust
emission detected in the molecular cloud zone (bars in the right panel, 
see details in \cite{ncr+03}).
\label{gc}}
\end{figure}

\vspace{-2mm}
\section{Magnetic fields in the Galactic center}

Within a few hundred pc of the Galactic center, both poloidal and toroidal
magnetic fields have been observed.
The non-thermal radio filaments (Fig.~\ref{gc}) discovered within
$1^{\circ}$ from the Galactic center (\cite{ymc84}, \cite{lme99},
\cite{yhc04}, \cite{nlk+04}, \cite{lnlk04}) indicate poloidal magnetic
fields in the region within a few hundred parsecs of the center of the
Galaxy. These filaments are highly polarized (\cite{lme99}) and almost
perpendicular to the Galactic plane, although some newly found
examples are not so perpendicular to the Galactic plane
(\cite{lnlk04}). RM studies show that the magnetic field is aligned
along the filaments (\cite{lme99}). The filaments are probably
illuminated flux tubes, with a field strength of about 1~mG
(\cite{ms96a}). LaRosa et al. (2005)\nocite{lbs+05} detected diffuse
radio emission of extent 400~pc and on this basis argued for a weak
pervasive field of tens of $\mu$G in the central region. However, this
is the volume-averaged field strength in such a large region. The
poloidal fields are possibly limited to a smaller central region.  The
newly discovered ``double helix'' nebula (\cite{mud06}), with an
estimated field strength of order 100~$\mu$G, reinforces the presence
of strong poloidal magnetic fields in tube format merging from the
rotating circumnuclear gas disk near the Galactic Center.

Polarized thermal dust emission has been detected in the molecular cloud
zone at sub-mm wavelength (see Fig.~\ref{gc}, \cite{ncr+03,cdd+03}), which
is probably related to the toroidal fields parallel to the Galactic plane
and complements the poloidal fields shown by the vertical filaments.  The
observed molecular cloud zone has a size of a few hundred pc, and is
possibly a ring-like cloud outside the central region with poloidal field
(see Fig.1 of \cite{cha01}). The sub-mm polarization observations of the
cloud zone offer information only about the field orientations. Zeeman
splitting measurements of HI absorption against Sgr~A (e.g. \cite{plc95}) or
of the OH maser in the Sgr A region (\cite{yrg+99}) give a line-of-sight
field strength of a few mG in the clouds. It is possible that toroidal
fields in the clouds are sheared from the poloidal fields, so that the RM
distribution of radio sources in this very central region could be
antisymmetric (\cite{ncr+03}).

Outside the central region of a few hundred pc to a few kpc, the
structure in the stellar and gas distributions and the magnetic
structure are all mysterious.  There probably is a bar. The
large-scale magnetic fields should be closely related to the material
structure but have not been revealed yet. The large positive RMs of
background radio sources within $|l|<6^{\circ}$ of the Galactic Center
(\cite{rrs05}) are probably related to magnetic fields following the
bar (\cite{rrs08}). Comparison of the RMs of these background radio
sources with RMs of foreground pulsars (see Fig.~\ref{rm_psr_egr})
should be helpful in delineating the field structure.

\begin{figure}[b]
\begin{minipage}[c]{0.6\linewidth}
\centering\includegraphics[width=45mm,angle=270]{ers.ps}
\end{minipage}%
\hspace{0.01\textwidth}
\begin{minipage}[c]{0.4\textwidth}
\centering\includegraphics[width=45mm,angle=0]{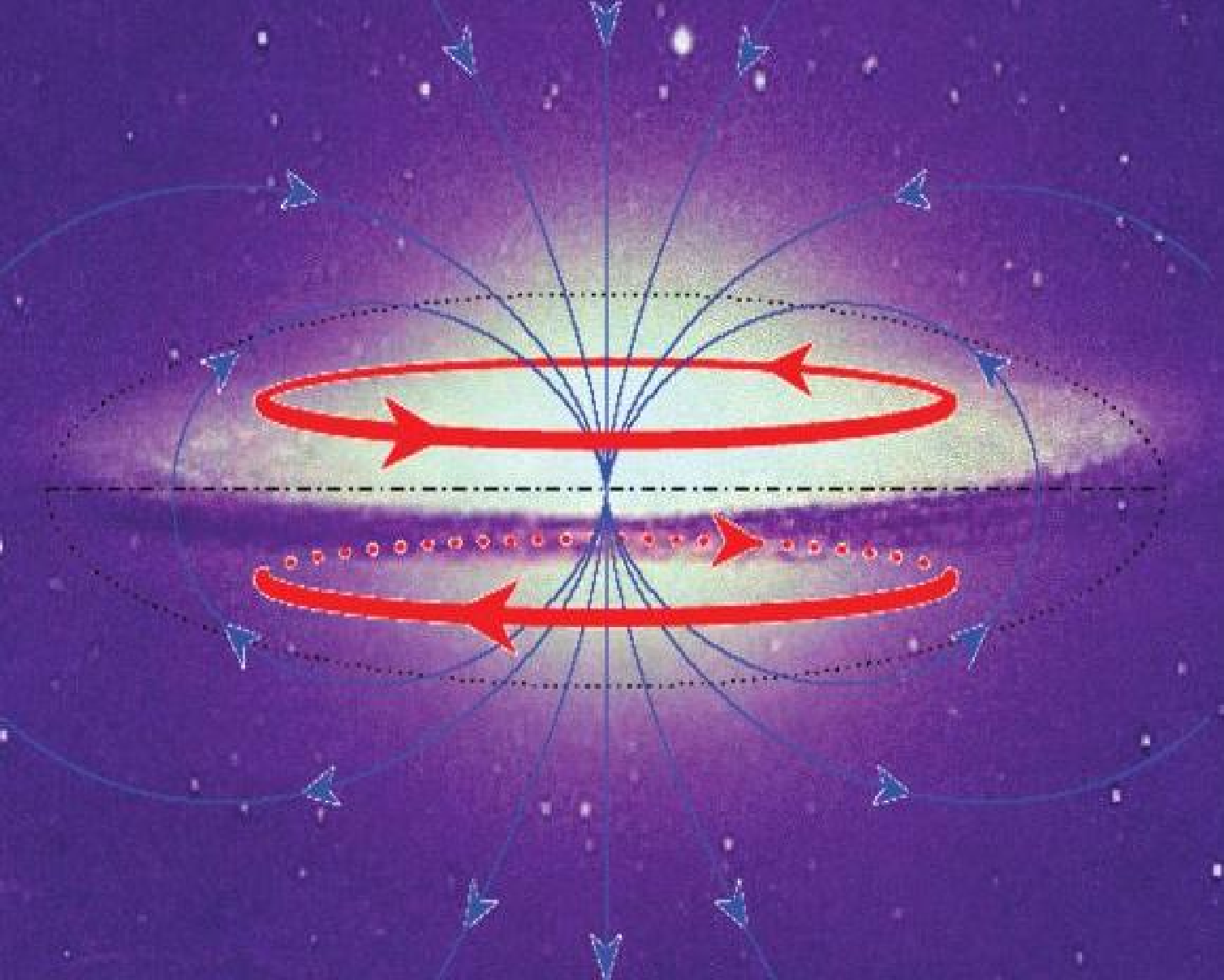}
\end{minipage}%
\caption{The antisymmetric rotation measure sky, derived from RMs of
extragalactic radio sources after filtering out the outliers with anomalous RM
values. The distribution corresponds to magnetic structure in the
Galactic halo as illustrated on the right. See Han et al. (1997, 1999).
\label{rmsky}}
\end{figure}

\vspace{-3mm}
\section{Magnetic fields in the Galactic halo}

Magnetic field structure in the Galactic halo can be revealed from RMs of
EGRs if allowance can be made for sources with outstanding intrinsic
RMs. The foreground Galactic RM is the common contribution to the observed
RMs of all EGRs within a small patch of sky. After ``anomalous'' RMs are
eliminated, the pattern for the Galactic RM can be obtained. Han et
al. (1997) \nocite{hmbb97} discarded any source if its RM deviated from the
average RM of neighbouring sources by more than 3$\sigma$, and obtained a
``cleaned'' RM sky.

A striking antisymmetry in the inner Galaxy with respect to Galactic
coordinates (see Fig.~\ref{rmsky}) has been identified from the cleaned RM
sky (\cite{hmbb97,hmq99}). The antisymmetry must result from azimuthal
magnetic fields in the Galactic halo with reversed field directions below
and above the Galactic plane. Such a field can be produced by an `A0' dynamo
mode. The observed filaments near the Galactic Center should result from the
dipole field in this dynamo scenario. The local vertical field component $B_z
\sim 0.2 \mu$G (\cite{hq94,hmq99}) may be a part of this dipole field in the
solar vicinity.

Han (2004) has shown that the RM amplitudes of extragalactic radio
sources in the mid-latitudes of the inner Galaxy are systematically
larger than those of pulsars, indicating that the antisymmetric
magnetic fields are not local but are extended towards the Galactic
center, far beyond the pulsars.  Model-fitting by Sun et
al. (2008)\nocite{srwe08} has confirmed this conclusion. The
antisymmetry has been shown more clearly after adding newly observed RMs of
more mid- or high latitude EGRs.

\vspace{-4mm}
\section{Concluding remarks}

Only after the magnetic field, $\vec{B}(r,\theta,z)$, is known at all
positions in our Galaxy will we have a complete picture of the
magnetic structure of the Milky Way. With only a small number of
measurements we must rely on fitting models to the data. ``Partial''
measurements in a greater number of regions, including quite large
regions, are now available and we are able to ``connect'' the
available measurements (pulsar RMs, maps of the RM sky) and outline
some basic features of the Galactic magnetic field.

However, as seen from the above observational review, many regions of our
Galaxy have not been observed very well for magnetic fields. For example,
the farther half of the Galactic disk and the RM sky are not well
known. Data in these regions are still scarce, either because of lack of
field probes or limited capability of current instruments. In addition, when
we try to get the large-scale field structure, the small-scale ``random''
fields, which are equally strong or even stronger, ``interfere'' with our
measurements to a large extent. In the first Galactic quadrant, we obtained
a large sample of pulsar RMs recently for the Galactic magnetic fields, but
the spiral structure and pulsar distances are very uncertain. Statistical
properties of fields on the energy-injection scales are not yet available,
but are crucial for theoretical studies of magnetic field origin and
maintenance. Magnetic fields in the disk-halo interface regions are
certainly complex but few observations are available. To fully observe and
understand the Galactic magnetic fields, there is a long way to go.

\vspace{-2mm}
\begin{acknowledgments}
Thanks to Drs. Tom Landecker and A. Nelson for reading the manuscript.
I am very grateful to colleagues who have collaborated with me for a long time
on work on Galactic magnetic fields:
Dr. R.N. Manchester from the Australia Telescope National Facility, CSIRO,
Prof. G.J. Qiao from Peking University (China),
%
%
%
%
Dr. Willem van Straten from Swinburne University (AU).
The author is supported by the National Natural Science Foundation (NNSF) of
China (10521001, 10773016 and 10833003) and the National Key Basic Research
Science Foundation of China (2007CB815403).
\end{acknowledgments}

\newpage
\normalsize
\noindent {\bf Questions and Answers}

\vspace{1mm}
\noindent{\bf A.H. Nelson:} Can we obtain any information on the field
  strength in the outer disk of our Galaxy from the extragalactic
  rotation measures, say from 20-30kpc radius?

\noindent{\bf J.L. Han:} Hard. From pulsar RMs, we can determine the
field strength up to $R\sim10$ kpc. For the outer Galaxy, the RMs of
extragalactic radio sources have to be fitted with a model of $B$ and
a model of $n_e$.  Given the many uncertainties in these models for
the outer disk, it is really hard to determine the field
strength. Nevertheless, from the determined variation of the field
strength with the Galactocentric radius, you can extrapolate the field
strength to the outer Galaxy.

\vspace{1mm}
\noindent{\bf Elisabete M. de Gouveia Dal Pino:} Do you see any
indication of an X-shaped magnetic structure similar to the one we
see in other edge-on star-forming galaxies?

\noindent{\bf J.L. Han:} For this purpose, one needs to map the
polarization in a large field-of-view around the Galactic center. At
present, the available good maps with a field-of-view of 1 or 2 degrees
were made with the VLA, and show the non-thermal filaments
indicating the poloidal fields.  It is impossible to restore the
absolute level of $U$ and $Q$ in these maps. Therefore we can
not map it now even if there may be such an X-shaped field
structure. It is possible that the double helix nebula seen by Morris
et al. (2006) and the filaments observed in the Galactic center region
are somehow the foot or leg part of such a structure.

\vspace{1mm}
\noindent{\bf T. Landecker:} I am concerned that the interpretation of
RMs of pulsar and extragalactic radio sources is heavily dependent on
the models of electron density distribution of Taylor \& Cordes (1993)
or Cordes \& Lazio (2002). These models incorporate implicitly a spiral
pattern for the Galaxy, going back to the work of Georgelin \& Georgelin
(1976) whose basic assumption is that there is a neat spiral pattern in
the structure of the Milky Way. Please comment.

\noindent{\bf J.L. Han:} You are right on the electron density
model. For interpretation for pulsar RMs, we work on the RM variations
against not only the pulsar distances but also the observed
DM. Considering that the relative positions of pulsars are more
closely related to the DM rather than the electron density model, we
believe that the magnetic structure derived from pulsar RMs and DMs
are less model-dependent. For interpretation for RMs of extragalactic
radio sources, yes, it is very heavily dependent on the electron
density model and non ``DM'' to use.

From all observations available, evidence for spiral arms is so strong,
therefore it is a correct idea to incorporate the spiral structure into any
electron density model. Now, the spiral arms in the fourth Galactic quadrant
have been better determined, while in the first quadrant, it is much less
clear. Also, the electron density within a few kpc from the Galactic center
is not clear at all. The electron density model and hence the pulsar
distances therefore have a large uncertainty in these regions. Future
electron density models will be probably improved there.

\begin{figure}[b]
\vspace{-3mm}
\centering
\includegraphics[width=55mm]{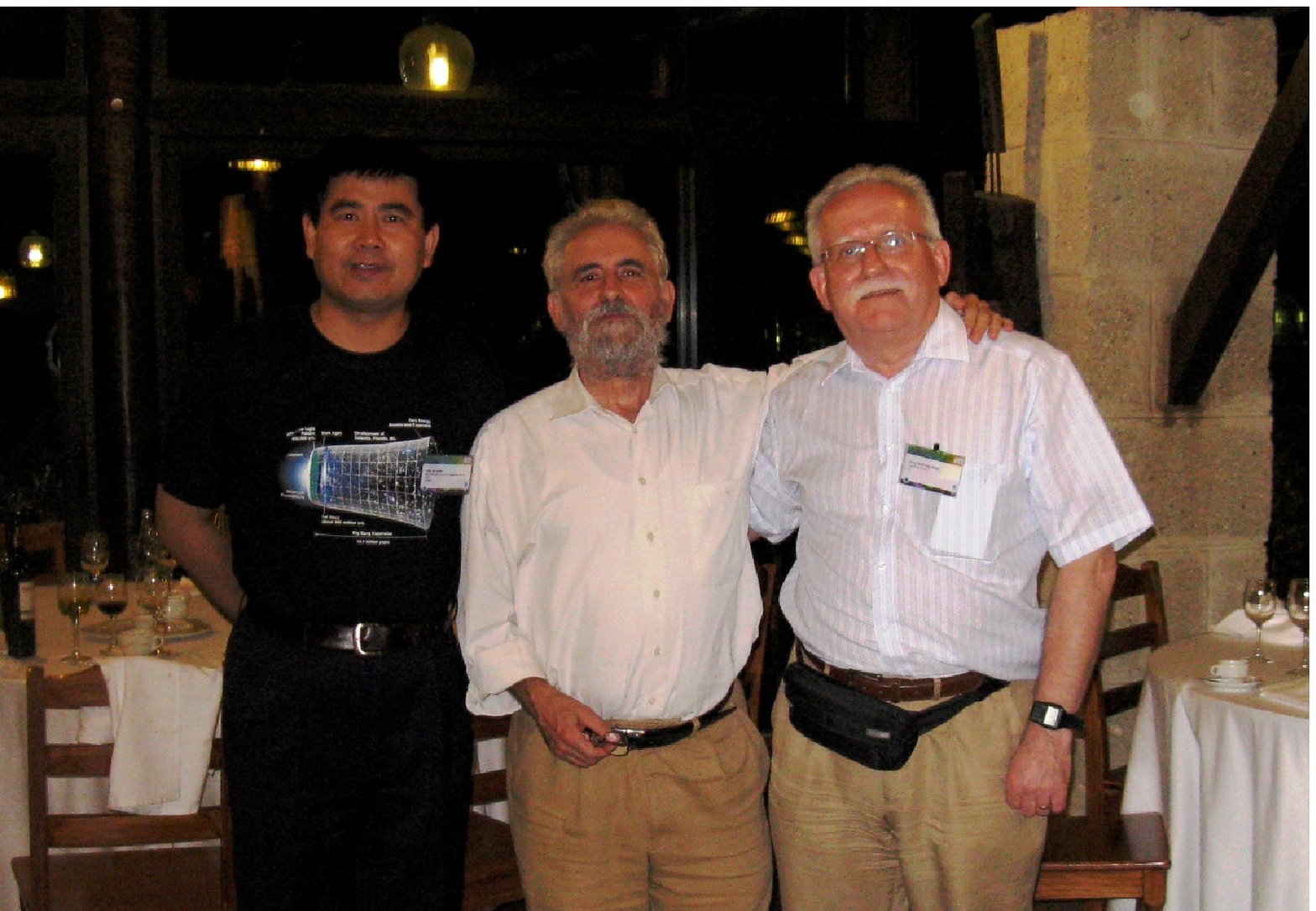}
\includegraphics[width=56mm]{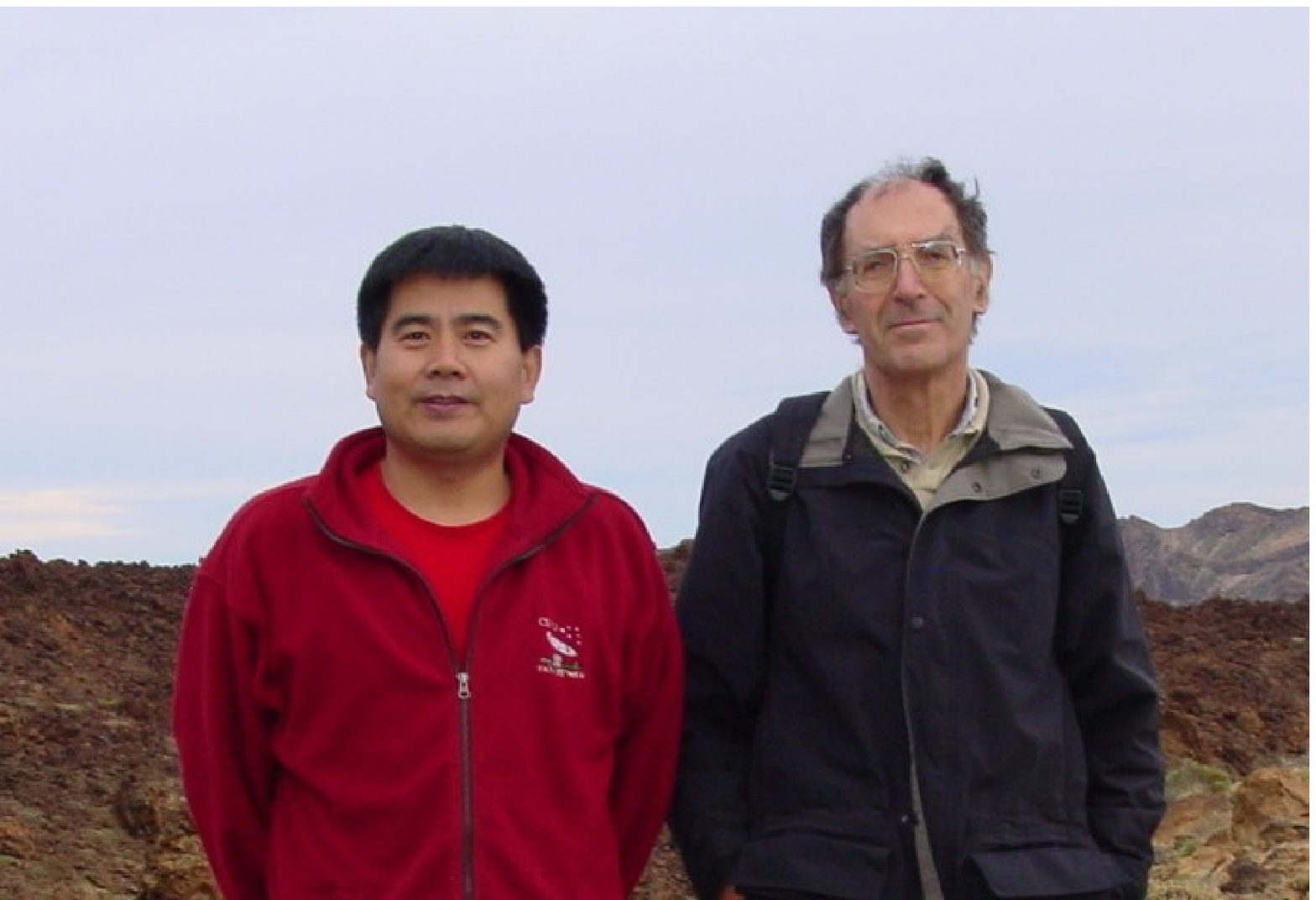}

\hspace{1mm}J.~L. Han, E. Battaner, and A. Nelson \hspace{8mm} J.L. Han and
T.  Landecker
\hspace{8mm}

\vspace{-3mm}
\end{figure}

\end{document}